\documentclass[aps,showpacs,nofootinbib,prd,twocolumn]{revtex4-1}
\usepackage{graphicx}
\usepackage{amssymb}
\usepackage{amsmath}
\usepackage{xcolor}
\usepackage{mathtools,slashed}
\usepackage{epstopdf}
\graphicspath{{./}}

\newcommand{\be}{\begin{equation}}
\newcommand{\ee}{\end{equation}}
\newcommand{\bea}{\begin{eqnarray}}
\newcommand{\eea}{\end{eqnarray}}
\newcommand{\beas}{\begin{eqnarray*}}
\newcommand{\eeas}{\end{eqnarray*}}

\newcommand{\slshh}[1]{{\not \!\! #1}}

\begin{document}
\title{Relaxation time for quark spin and thermal vorticity alignment in heavy-ion collisions}
\author{Alejandro Ayala$^{1,2}$, David de la Cruz$^{1,3}$, S. Hern\'andez-Ort\'iz$^1$, L. A. Hern\'andez$^{1,2}$  and Jordi Salinas$^1$}
  \address{
  $^1$Instituto de Ciencias
  Nucleares, Universidad Nacional Aut\'onoma de M\'exico, Apartado
  Postal 70-543, Ciudad de M\'exico 04510,
  M\'exico\\
  $^2$Centre for Theoretical and Mathematical Physics, and Department of Physics,
  University of Cape Town, Rondebosch 7700, South Africa\\
  $^3$Departamento de F\'isica, Escuela Superior de F\'isica y Matem\'aticas
del Instituto Polit\'ecnico Nacional,
Unidad Adolfo L\'opez Mateos, Edificio 9, 07738 Ciudad de M\'exico, M\'exico\\}

\begin{abstract}

We compute the relaxation time for quark/antiquark spin and thermal vorticity alignment in a quark-gluon plasma at finite temperature and quark chemical potential. We model the interaction of quark/antiquark spin with thermal vorticity as driven by a phenomenological modification of the elementary quark interaction with gluons. We find that in a scenario where the angular velocity of the quark-gluon plasma produced in a peripheral heavy-ion collision is small, quarks/antiquarks take a long time to align their spin with the vorticity. However, when the angular velocity created in the reaction is large, the alignment is efficient and well within the lifetime of the system created in the reaction. The relaxation time is larger for antiquarks which points out to a difference for the polarization of hadrons and antihadrons when this alignment is preserved during hadronization.

\end{abstract}


\maketitle

Collisions of heavy nuclei at high energy produce locally equilibrated matter whose properties have been successfully described in terms of concepts and techniques borrowed from hydrodynamics. One of such concepts is the {\it thermal vorticity}~\cite{Becattini2015} defined as
\bea
   \overline{\omega}_{\mu\nu}=\frac{1}{2}\left(\partial_\nu\beta_\mu - \partial_\mu\beta_\nu\right),
\label{thervor}
\eea
where $\beta_\mu=u_\mu(x)/T(x)$, with $u_\mu(x)$ the local fluid four-velocity and $T(x)$ the local temperature. Thermal vorticity can be produced in peripheral collisions where the colliding matter develops an orbital angular momentum, and thus an angular velocity $\vec{\omega}=\omega\hat{z}$, normal to the reaction plane, hereafter chosen as the direction of the $\hat{z}$ axis. The orbital angular momentum is due to the inhomogeneity of the matter density profile in the transverse plane~\cite{Becattini2008}. For a constant angular velocity and uniform  temperature, the magnitude of the thermal vorticity is given by $\omega/T$.

The possibility to develop a local alignment of particle spin along the thermal vorticity, has prompted the search for consequences, among them, the chiral vortical effect~\cite{chiralvortef} and the global polarization of hadrons, most notably of hyperons~\cite{Becattini2017,Csernai,Sorin2016,Sorin2017,Xie,nature,STARorig,Pang,Sun, Han,Xia,Teryaev,Karpenko,Suvarieva,Kolomeitsev,Xie2,Guo,Ma}. Moreover, recent measurements of different global polarization of $\Lambda$ and $\overline{\Lambda}$, as the collision energy decreases~\cite{STAR}, motivate the need for a deeper understanding of the conditions for relaxation between angular momentum and spin degrees of freedom and of its dependence on the collision parameters such as energy, impact parameter, temperature and quark chemical potential.

Theoretical studies that address these consequences typically assume that such alignment does occur. However, to our knowledge, it has been only recently that an estimate of the relaxation time for the strange quark spin and vorticity alignment has been performed in Ref.~\cite{Kapusta}. This work has resorted to study the alignment of the strange quark spin induced either by vorticity fluctuations or helicity flip from interactions with light quarks and gluons, finding that within these mechanisms, the obtained relaxation time is too large.

In this work, we address, from a thermal field-theoretical point of view, the question of whether or not the transferring of angular momentum to spin degrees of freedom is fast enough such that searches for global particle polarization in relativistic heavy-ion collisions can be put on firmer grounds. Our strategy is to compute the relaxation time for the interaction of thermal vorticity and quark/antiquark spin driven by a phenomenological modification of the elementary interaction between quarks and gluons, accounting not only for temperature but also for quark chemical potential effects. Since the relaxation time turns out to be inversely proportional to the magnitude of the vorticity, we provide estimates  using values obtained in scenarios where either a small or a large fraction of the angular momentum imparted on the participants is preserved. 

Consider a QCD plasma in thermal equilibrium at temperature $T$ and quark chemical potential $\mu$. The interaction rate $\Gamma$ of a quark with four-momentum $P=(p_0,\vec{p})$ can be conveniently expressed in terms of the quark  self-energy $\Sigma$ as
\bea
\Gamma(p_0)=\tilde{f}(p_0){\mbox{Tr}}\left[\gamma^0{\mbox{Im}}\Sigma\right],
\label{rate}
\eea
where $\tilde{f}(p_0)$ is the Fermi-Dirac distribution.

\begin{figure}[t]
 \begin{center}
  \includegraphics[scale=0.65]{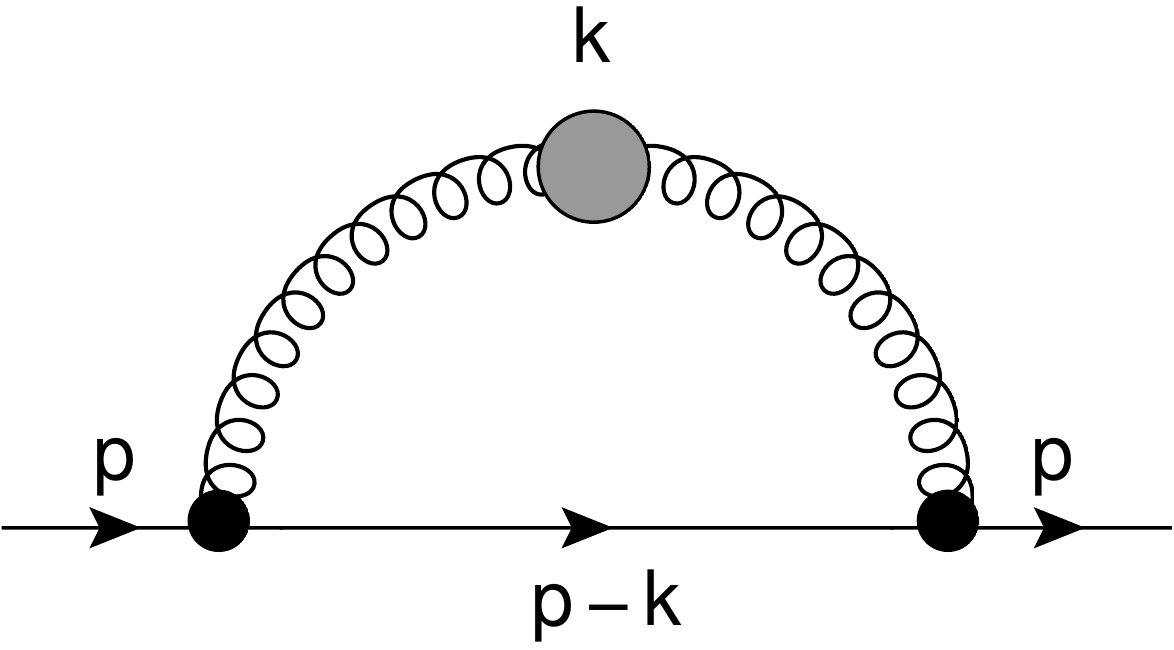}
 \end{center}
 \caption{One-loop quark self-energy diagram that also serves to define the kinematics. The gluon line with a blob represents the effective gluon propagator at finite density and temperature in the HTL approximation. The blobs on the quark-gluon vertices represent the effective coupling between the quark spin and the vorticity.}
 \label{Fig1}
\end{figure}

The one-loop contribution to $\Sigma$, depicted in Fig.~\ref{Fig1}, is given explicitly by
\be
   \Sigma=T\sum_n\int\frac{d^3k}{(2\pi)^3}\lambda^\mu_a S(\slshh{P}-\slshh{K})\lambda^\nu_b {}^*G_{\mu\nu}^{ab}(K)\ ,
   \label{sigma}
\ee
where $S$ and $^*G$ are the quark and effective gluon propagators, respectively. Using the imaginary-time formalism of thermal field theory, the incoming quark and virtual gluon four-momenta become $P=(i\tilde{\omega}_m + \mu,\vec{p})$ and $K=(i\omega_n,\vec{k})$, respectively, with $\tilde{\omega}_m=(2m+1)\pi T$ and $\omega_n=2n\pi T$, $m$ and $n$ being integers. In order to introduce the interaction between the thermal vorticity and the quark spin, we consider an effective vertex of the form  
\bea
    \lambda^\mu_a = g \frac{\sigma^{\alpha\beta}}{2}\overline{\omega}_{\alpha\beta}\gamma^\mu t_a,
    \label{vertex}
\eea
where $\sigma^{\alpha\beta}/2$, with $\sigma^{\alpha\beta}=\frac{i}{2}\left[\gamma^\alpha,\gamma^\beta\right]$ is the quark spin operator and $t_a$ are the color matrices in the fundamental representation. This vertex models the alignment between  quark spin and thermal vorticity driven by the elementary quark interaction with gluons in QCD.

The effective gluon propagator is obtained by summing the geometric series for the one-loop gluon polarization tensor at high temperature and/or quark chemical potential. The intermediate quark line is taken as a bare quark propagator such that the inverse of the interaction rate corresponds to the relaxation time for the spin and vorticity alignment for quarks that are originally not thermalized. For simplicity, we work in the approximation where the quark mass vanishes.

In a covariant gauge, the Hard Thermal Loop (HTL) approximation to the effective gluon propagator is given by
\bea
   ^*G_{\mu\nu}(K)=\! ^*\!\Delta_L(K)P_{L\, \mu\nu} +\, ^*\!\Delta_T(K)P_{T\, \mu\nu},
\label{effectiveglue}
\eea
where $P_{L,T\, \mu\nu}$ are the polarization tensors for three-dimensional longitudinal and transverse gluons, both of which are, of course, four-dimensionally transverse. The gluon propagator functions for longitudinal and transverse modes, $^*\!\Delta_{L,T}(K)$, are given by 
\bea
^*\!\Delta_{L}(K)^{-1}&=&K^2+2m^2\frac{K^2}{k^2}\left[1-\left(\frac{i\omega_n}{k}\right)Q_0\left(\frac{i\omega_n}{k}\right)\right]\nonumber\\
^*\!\Delta_{T}(K)^{-1}&=&-K^2-m^2\left(\frac{i\omega_n}{k}\right)\left[\left[1-\left(\frac{i\omega_n}{k}\right)^2\right]\right.\nonumber\\
&\times&
Q_0\left(\frac{i\omega_n}{k}\right)+\left.\left(\frac{i\omega_n}{k}\right)\right],
\label{scalarfunc}
\eea
where
\bea
Q_0(x)=\frac{1}{2}\ln\left(\frac{x+1}{x-1}\right),
\label{Legendre}
\eea
and $m$ is the gluon thermal mass given by
\bea
m^2=\frac{1}{6}g^2C_AT^2+\frac{1}{12}g^2C_F\left(T^2+\frac{3}{\pi^2}\mu^2\right),
\label{gluonmass}
\eea
where $C_A=3$ and $C_F=4/3$ are the Casimir factors for the adjoint and fundamental representations of $SU(3)$, respectively~\cite{LeBellac}.

The sum over Matsubara frequencies involves products of the propagator functions for longitudinal and transverse gluons, $^*\Delta_{L,T}$ and the Matsubara propagator for the bare quark $\tilde{\Delta}_F$, such that the term that depends on the summation index can be expressed as
\bea
S_{L,T}=T\sum_n\, ^*\Delta_{L,T}(i\omega_n)\tilde{\Delta}_F(i(\omega_m-\omega_n)).
\label{sumprod}
\eea
This sum is more straightforward evaluated introducing the spectral densities $\rho_{L,T}$ and $\tilde{\rho}$ for the gluon and fermion propagators, respectively. The imaginary part of $S$ can thus be written as
\bea
   {\mbox{Im}}\,S_{L,T}&=&\pi\left(e^{(p_0-\mu)/T}+1\right)\int_{-\infty}^{\infty}\frac{dk_0}{2\pi}\int_{-\infty}^{\infty}\frac{dp_0'}{2\pi}f(k_0)\nonumber\\
   &\times&\tilde{f}(p_0'-\mu)\delta(p_0-k_0-p_0')\rho_{L,T}(k_0)\tilde{\rho}(p_0'),
\label{Imsum}
\eea
where $f(k_0)$ is the Bose-Einstein distribution. The spectral densities $\rho_{L,T}(k_0,k)$ are obtained from the imaginary part of $^*\Delta_{L,T}(i\omega_n,k)$ after the analytic continuation $i\omega_n\to k_0+i\epsilon$ and contain the discontinuities of the photon propagator across the real $k_0$-axis. Their support depends on the ratio $x=k_0/k$. For $|x|>1$, $\rho_{L,T}$ have support on the (time-like) quasiparticle poles. For $|x|<1$ their support coincides with the branch cut of $Q_0(x)$. On the other hand, the spectral density corresponding to a bare quark is given by
\bea
\tilde{\rho}(p_0')=2\pi\epsilon(p_0')\delta({p_0'}^2-E_p^2),
\label{spectralquark}
\eea
where $E_p=|\vec{p}-\vec{k}|$. The kinematical restrictions that Eq.~(\ref{spectralquark}) imposes on Eq.~(\ref{Imsum}), limit the integration over gluon energies to the space-like region, namely, $|x|<1$, therefore, the part of the gluon spectral densities that contribute to the interaction rate are given by
\begin{widetext}
\bea
\rho_L(k_0,k)&=&\frac{x}{1-x^2}\frac{2\pi m^2\theta(k^2-k_0^2)}{\left[k^2+2m^2\left(1-(x/2)\ln |(1+x)/(1-x)|\right)\right]^2+\left[\pi m^2x\right]^2}
\nonumber\\
\rho_T(k_0,k)&=&\frac{\pi m^2x(1-x^2)\theta(k^2-k_0^2)}{\left[k^2(1-x^2)+m^2\left(x^2+(x/2)(1-x^2)\ln |(1+x)/(1-x)|\right)\right]^2+\left[(\pi/2)m^2x(1-x^2)\right]^2}.
\label{specdenglu}
\eea
Collecting all the ingredients, the interaction rate for a quark with energy $p_0$ to align its spin with the thermal vorticity is given by
\bea
\Gamma\left(p_0\right)&=&\frac{\alpha_s}{4\pi}\left(\frac{\omega}{T}\right)^2\frac{C_F}{p_0}\int_0^\infty\!\!\! dk\ k \int_{-k}^k dk_0\theta(2p_0-k+k_0)
(1+f(k_0))\tilde{f}(p_0+k_0-\mu)
\sum_{i=L,T}C_i(p_0,k_0,k)\rho_i(k_0,k),
\label{Gammapart}
\eea
\end{widetext}
where the functions $C_{L,T}$ come from the contraction of the polarization tensors $P_{L,T\, \mu\nu}$ with the trace of the factors involving gamma matrices in Eq.~(\ref{rate}) with $\Sigma$ given by Eq.~(\ref{sigma}). For consistency of the approximation where we have considered massless quarks, we have also dropped terms proportional to the quark four-momentum components. After implementing the kinematical restriction for the allowed values of the angle between the quark and gluon momenta, these functions are given explicitly by
\bea
\!\!\!\!\!\!C_T(p_0,k_0,k)&=&8k_0\left(\frac{k^2-2k_0 p_0-k_0^2}{2k p_0}\right)^2\nonumber\\
\!\!\!\!\!\!C_L(p_0,k_0,k)&=&-8k_0\left[\left(\frac{k^2-2k_0 p_0-k_0^2}{2k p_0}\right)^2-\frac{1}{2}\right]\!.
    \label{Cs}
\eea
\begin{figure}[b]
 \begin{center}
  \includegraphics[scale=0.44]{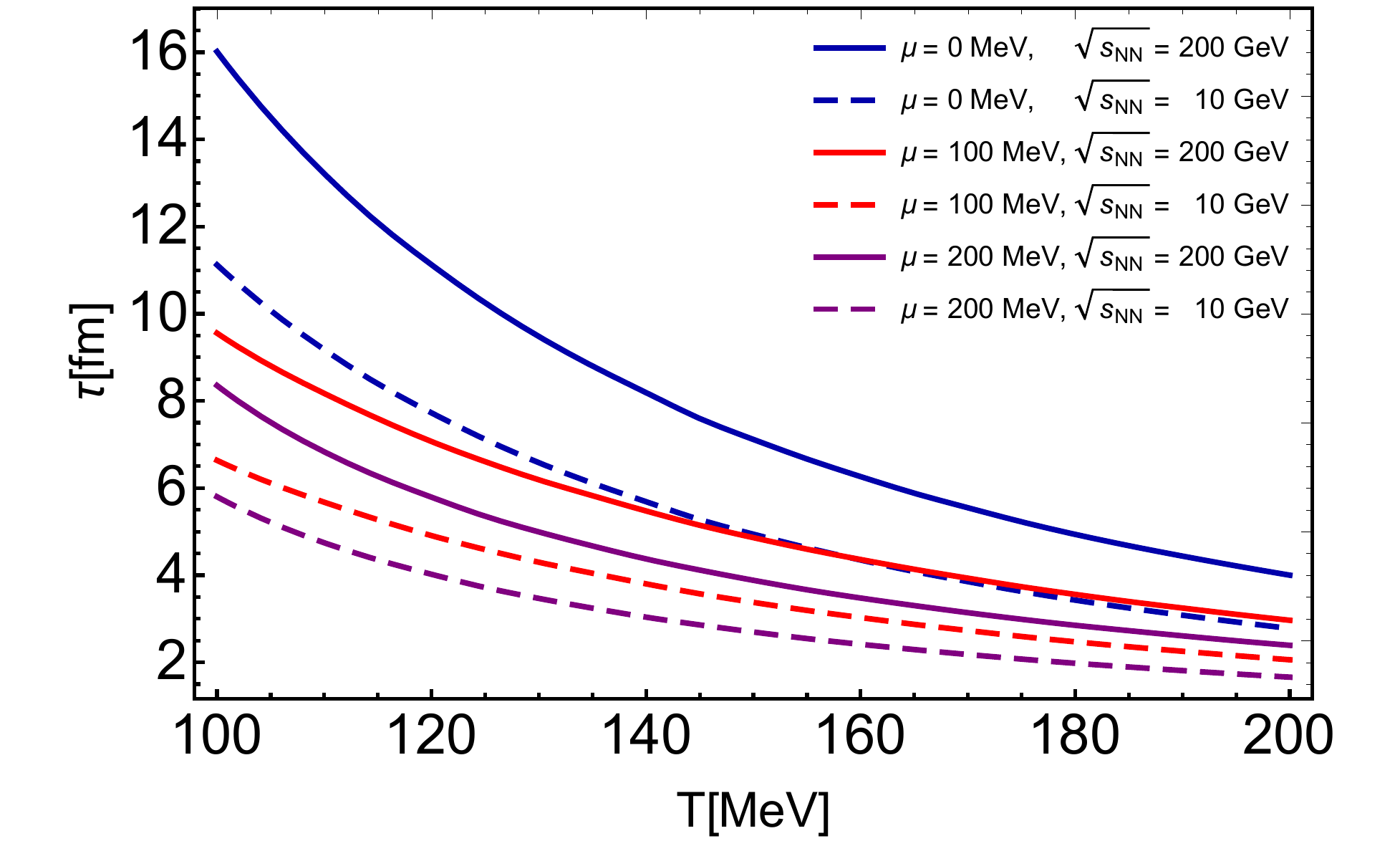}
 \end{center}
 \caption{Relaxation time $\tau$ for quarks as a function of temperature for semicentral collisions at a fixed impact parameter $b=10$ fm for two collision energies, $\sqrt{s_{NN}}=10, 200$ GeV, for which, using the findings of Ref.~\cite{Deng,Jiang}, $\omega\simeq0.12, 0.10$ fm$^{-1}$, respectively.  Notice that $\tau$ is of order $\lesssim$ 3 fm only for the largest $T$ and $\mu$ considered.}
\label{fig2}
\end{figure}
The total interaction rate is obtained by integrating Eq.~(\ref{Gammapart}) over the available phase space
\bea
\Gamma=V \int \frac{d^3p}{(2\pi)^3}\Gamma(p_0),
\label{Gammatot}
\eea
where $V$ is the volume of the overlap region in the collision and for massless quarks $p_0=p$. Recall that, for the collision of symmetric systems of nuclei with radii $R$ and a given impact parameter $b$, is given by
\bea
   V=\frac{\pi}{3}(4R+b)(R-b/2)^2.
\label{volume}
\eea

Putting all these ingredients together, we use the expression for $\Gamma$ from Eq.~(\ref{Gammatot}) to study the parametric dependence of the relaxation time for spin and vorticity alignment, defined as
\bea
\tau \equiv 1/\Gamma.
\label{relax}
\eea
For the analysis, hereafter we use the conservative value $\alpha_s= 0.3$. We consider Au+Au collisions ($R=7.27$ fm) 
at $\sqrt{s_{NN}}$=10, 200 GeV
for semicentral collisions with impact parameter of $b=10$ fm, where the maximum angular momentum is expected to be imparted.

\begin{figure}[b]
 \begin{center}
  \includegraphics[scale=0.44]{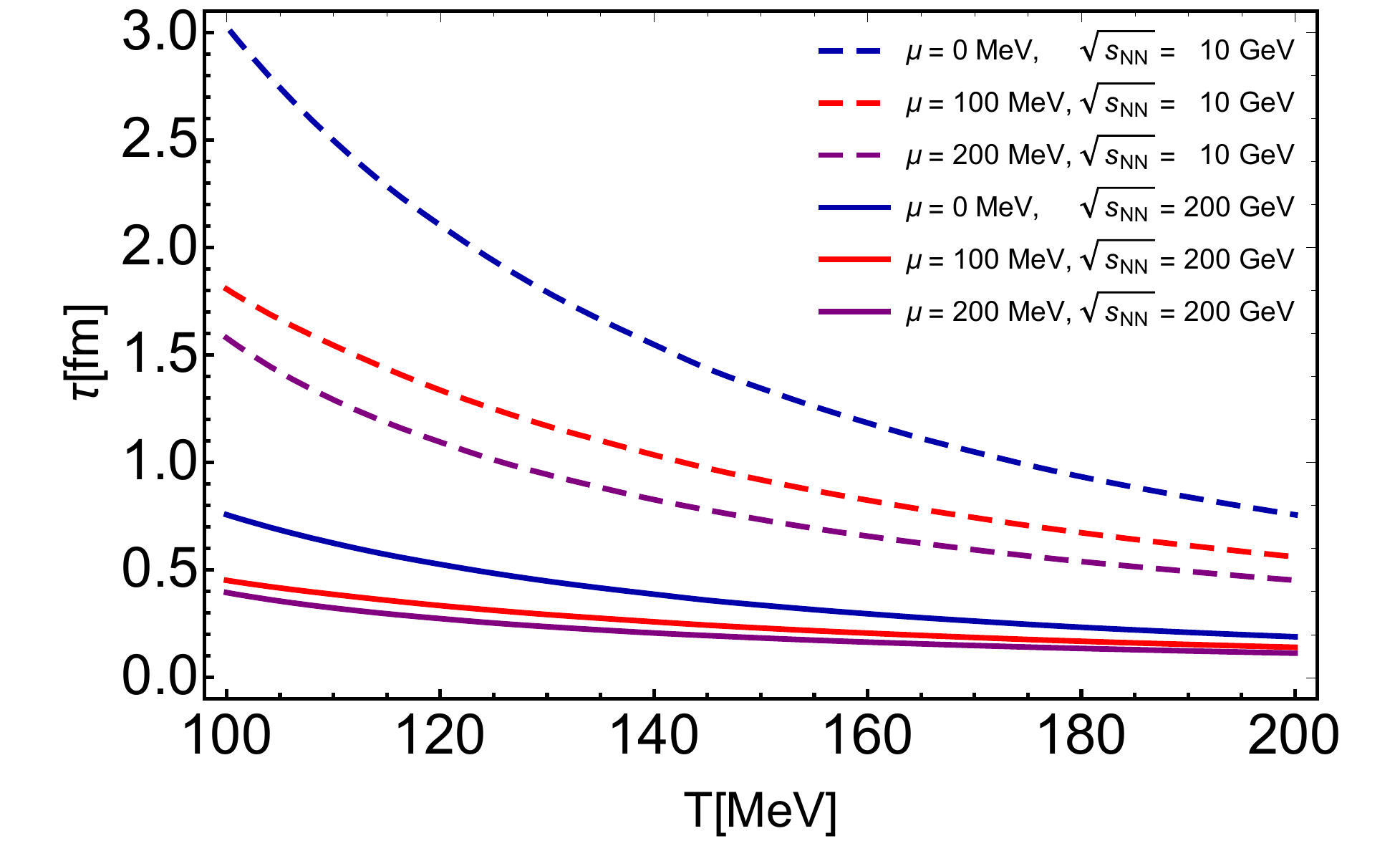}
 \end{center}
 \caption{Relaxation time $\tau$ for quarks as a function of temperature for semicentral collisions at a fixed impact parameter $b=10$ fm for two collision energies, $\sqrt{s_{NN}}=10, 200$ GeV, for which, using UrQMD simulations~\cite{UrQMD} $\omega=0.23, 0.46$ fm$^{-1}$, respectively. Notice that $\tau \lesssim 3$ fm for all $T$ and $\mu$ considered.}
\label{fig3}
\end{figure}
To estimate the magnitude of the angular velocity, we consider two extreme situations.
First we take the results of Refs.~\cite{Deng, Jiang} where a full calculation of vorticity is performed using Hijing and AMPT. The magnitude of $\omega$ is computed at a time $\Delta t=0.4$ fm after full nuclei overlap. Vorticity of the QGP is small and the corresponding angular momentum is only a fraction of order 10\% of the angular momentum of the total
participants in the
interaction region. The estimated magnitude of $\omega$ is thus also small and found to slowly decrease with the collision energy from about $\omega\sim 0.12$ fm$^{-1}$ for $\sqrt{s_{NN}}\sim 10$ GeV to $\omega\sim 0.10$ fm$^{-1}$ for $\sqrt{s_{NN}}\sim 200$ GeV. 
Figure~\ref{fig2} shows the temperature dependence of the relaxation time for different values of the quark chemical potential. Notice that for the temperature range considered, $\tau\lesssim 3$ fm only for the largest $T$ and $\mu$ considered. This represents an indication that vorticity is transferred to quark spin degrees of freedom faster for larger values of $\mu$ and $T$ but for $\mu\sim 0$, the equilibration is not within the lifetime of the created system ($\simeq 10$ fm). 

\begin{figure}[t]
 \begin{center}
  \includegraphics[scale=0.44]{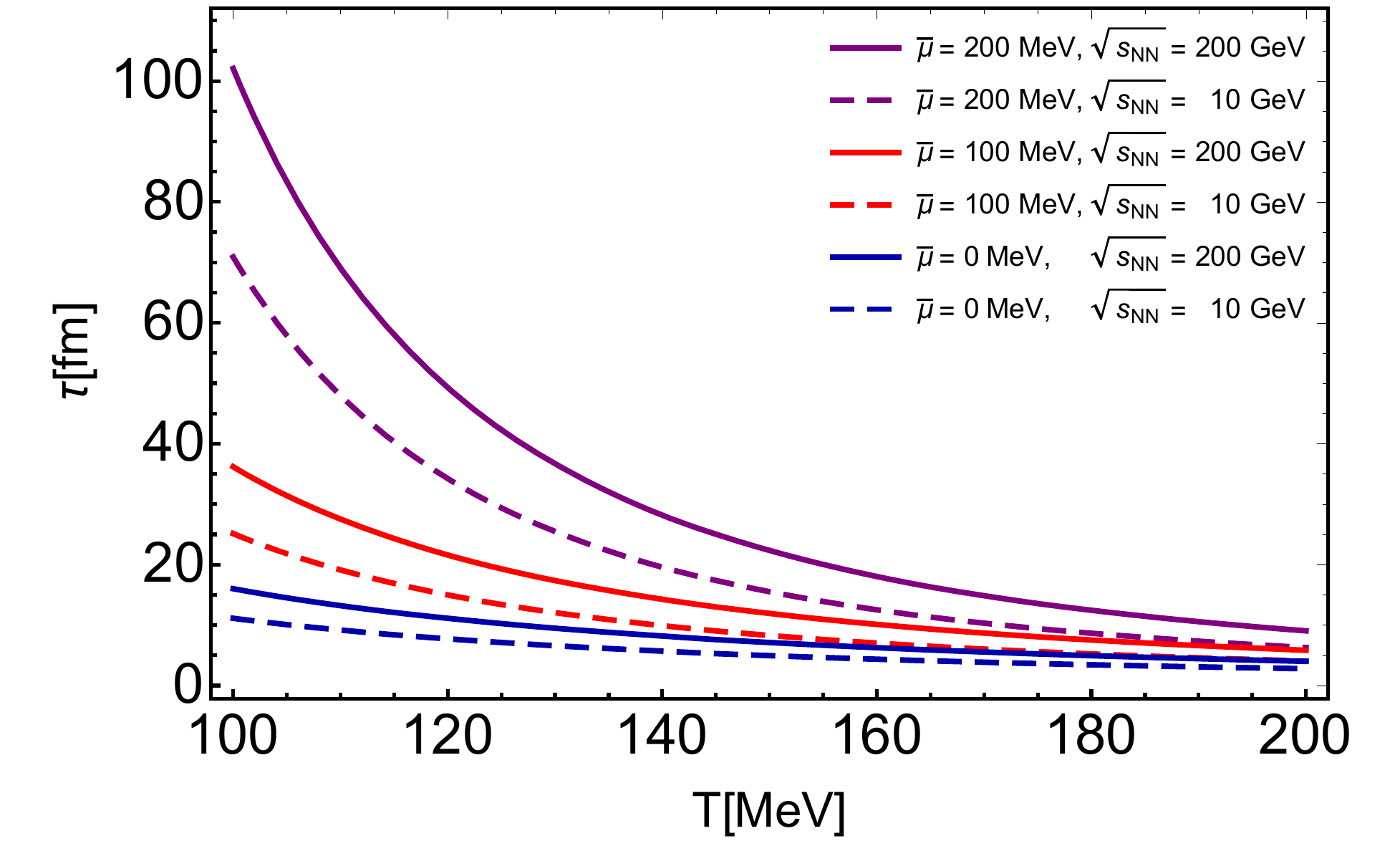}
 \end{center}
 \caption{Relaxation time $\tau$ for antiquarks as a function of temperature for semicentral collisions at a fixed impact parameter $b=10$ fm for two collision energies, $\sqrt{s_{NN}}=10, 200$ GeV, for which, using the findings of Ref.~\cite{Deng,Jiang}, $\omega\simeq0.12, 0.10$ fm$^{-1}$, respectively.  Notice that $\tau$ is of order $\lesssim$ 10 fm only for the largest $T$ and that it increases as $\bar{\mu}$ increases.}
°\label{fig4}
\end{figure}

Next, we consider a scenario where the total initial angular momentum of the participants is retained by the produced QGP. In this scenario we do not consider the competing effect of transverse expansion and $\omega$ turns out to be large. We perform UrQMD simulations~\cite{UrQMD} for Au+Au collisions. For both energies, $2\times10^4$ events were generated. The estimated angular velocity at a time $\Delta t=0.4$ fm also after full nuclei overlap, is computed non-relativistically as the average
\bea
   \omega=\frac{1}{N}\sum_{i=1}^Nv_i/r_i,
\label{average}
\eea
where $v_i$ and $r_i$ are the initial velocity along the beam axis and distance to the normal to the reaction plane that bisects the overlap region, for each particle that takes part of the reaction at the beginning of the collision, respectively. In this case, $\omega$ is found to increase with the collision energy from about $\omega\sim 0.23$ fm$^{-1}$ for $\sqrt{s_{NN}} = 10$ GeV to $\omega\sim 0.46$ fm$^{-1}$ for $\sqrt{s_{NN}} = 200$ GeV. Notice that Eq.~(\ref{average}) is only valid for  our UrQMD simulation which in turn is only used as a means to produce the initial participant particle velocity and position profiles. Because of this assumption, all of the energy goes into vorticity.

Notice that according to the analysis of Ref.~\cite{Deng}, the global angular momentum of the QGP manifests itself mainly in the form of local fluid shear rather than a global rigid rotation. Therefore, in this reference, $\Delta t = 0.4$ fm after full nuclei overlap is used as the time where such local fluid shear can start to be estimated. In order to compare the results of this reference to ours, for our UrQMD simulation we also set the same time after full nuclei overlap to estimate the angular velocity, which in our case does correspond to rigid rotation. This is of course an artificial way to describe the collision but it serves our purposes to estimate the relaxation time in an extreme favorable scenario.

Figure~\ref{fig3} shows the temperature dependence of the relaxation time for different values of the quark chemical potential. Notice that for the temperature range considered, $\tau\lesssim 3$ fm and decreases as $\mu$ increases. In this favorable scenario, with a larger magnitude of the angular velocity, vorticity is efficiently transferred to quark spin degrees of freedom, even for $\mu\sim 0$, and this transferring is even faster for larger values of $\mu$, for which we find that the relaxation time is well within the lifetime of the created system. 

\begin{figure}[t]
 \begin{center}
  \includegraphics[scale=0.44]{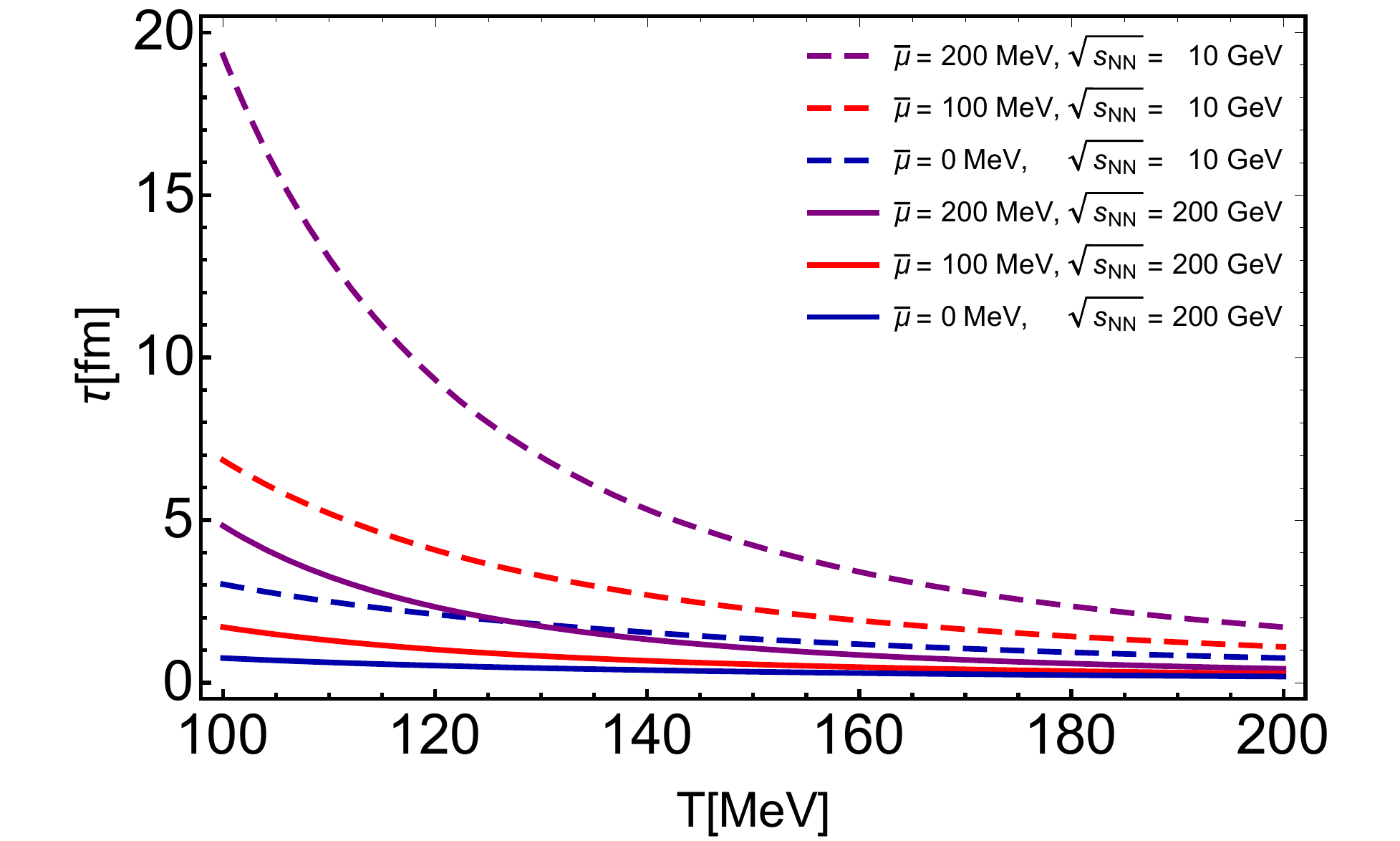}
 \end{center}
 \caption{Relaxation time $\tau$ for antiquarks as a function of temperature for semicentral collisions at a fixed impact parameter $b=10$ fm for two collision energies, $\sqrt{s_{NN}}=10, 200$ GeV, for which, using UrQMD simulations~\cite{UrQMD}, $\omega=0.23, 0.46$ fm$^{-1}$, respectively. Notice that $\tau$ is of order $\lesssim$ 10 fm for $T$ above 120 MeV, for all $\bar{\mu}$ considered.}
\label{fig5}
\end{figure}

In order to compute the corresponding relaxation times for antiquarks, it is only necessary to replace $\mu\to\bar{\mu}=-\mu$ in Eq.~(\ref{Imsum}). Figures~\ref{fig4} and~\ref{fig5} show the corresponding relaxation times for antiquarks obtained for the scenarios where $\omega$ is small and large, respectively. Notice that antiquarks take longer to align its spin to the vorticity as $\bar{\mu}$ increases and that for small $\omega$ the relaxation time is larger than the system's lifetime ($\simeq 10$ fm) except for the largest temperatures considered. For large $\omega$, vorticity and spin alignment happens faster and within the system's lifetime. The result is easy to understand by noticing that, from Eq.~(\ref{Gammapart}), $\Gamma$ is proportional to the quark/antiquark occupation number which becomes larger/smaller as $\mu$/$\bar{\mu}$ increases, which translates into a smaller/larger relaxation time. 

\begin{figure}[t!]
 \begin{center}
  \includegraphics[scale=0.43]{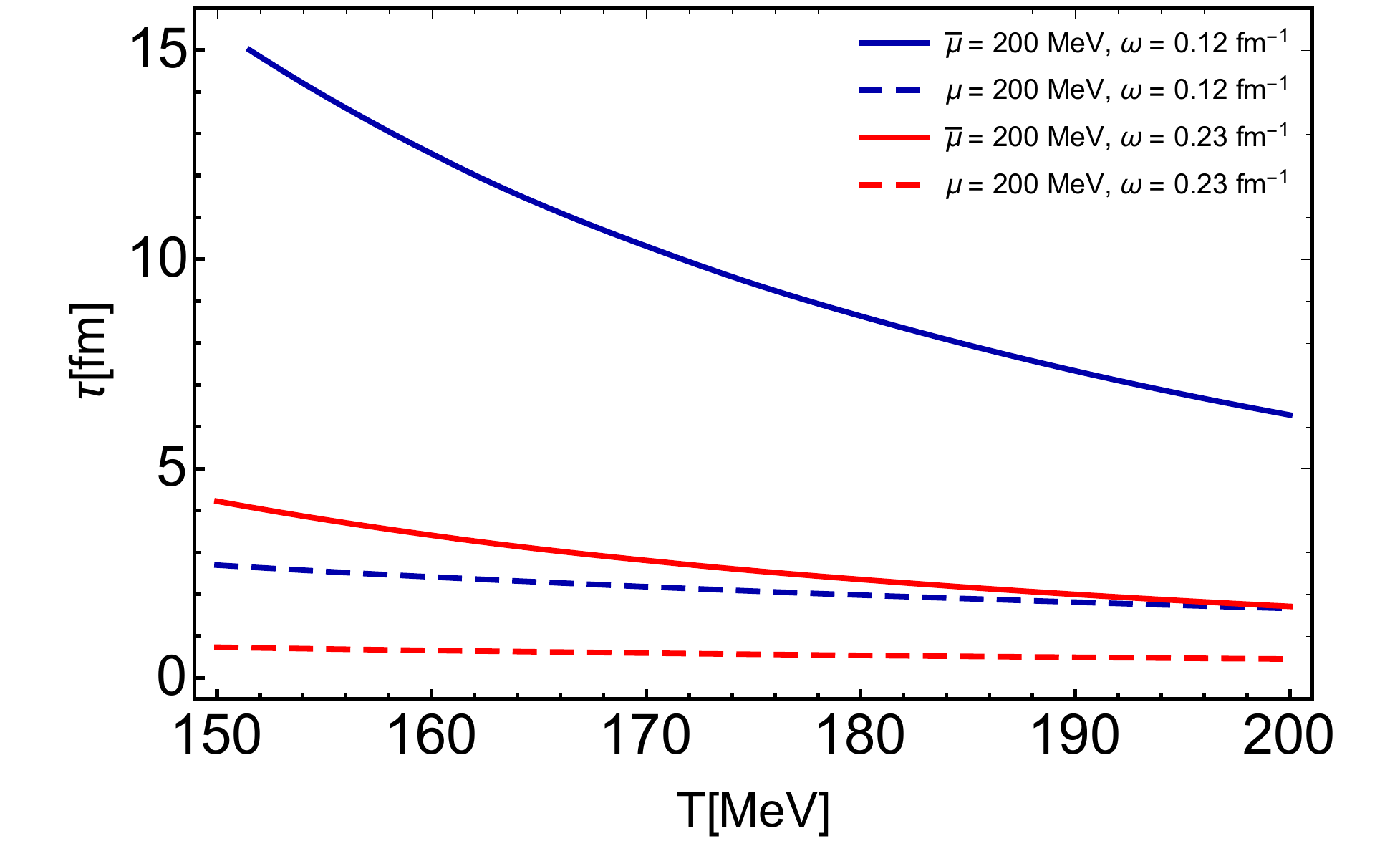}
 \end{center}
 \caption{Relaxation time $\tau$ for quarks and antiquarks as a function of temperature for semicentral collisions with $\sqrt{s_{NN}}=10$ GeV at a fixed impact parameter $b=10$ fm for small and large $\omega$ values and a fixed $\mu=200$ MeV. The behavior shows that in the former case, antiquark spin and vorticity align much slower than in the latter.}
\label{fig6}
\end{figure}

In order to contrast directly the relaxation times for quarks and antiquarks, for the scenarios where $\omega$ is large or small, Fig.~\ref{fig6} shows the corresponding $\tau$ as a function of $T$ for a range of temperatures above the expected phase transition temperature $150$ MeV $<T< 200$ MeV for $\mu$ close to the values expected to be achieved at NICA energies~\cite{NICA}. In order to understand why the relaxation time decreases with temperature, one has to recall that in equilibrium, particle reaction rates increase for larger temperatures. This is due to the increase of the phase space available when temperature increases, since the equilibrium distribution function allows for particles with larger energies to contribute to the reaction. Since the relaxation time is the inverse of the interaction rate, the former decreases as the temperature increases.

In conclusion, we have introduced a phenomenological description to study the temperature and quark chemical potential dependence of the relaxation time for quark spin and vorticity to align. Since vorticity is a global concept, its microscopic description requires modeling. We have used Eq.~(\ref{vertex}) to model such alignment. Although other modelings are possible, they should all contain features such as being proportional to the medium's angular velocity, the spin and to the strong coupling constant. A simple choice like the one made in this work already captures the essence of the effect.

Recall that in strange quark matter made up of massless free quarks, charge neutrality and beta equilibrium imply that the number of light quark flavors $u,d,s$ and their corresponding quark chemical potentials are the same. Therefore, our findings apply equally to the case where $\mu$ refers either to $u$, $d$ or $s$ chemical potentials. In order to make more quantitative predictions, to discriminate between the results for light flavors  and the strange quark, it is important to include the quark mass into these considerations. It is also important to consider the effects on the magnitude of vorticity coming from quark chemical potential~\cite{Saha}. This calculation is being prepared and will be reported elsewhere. Nevertheless, while the transferring of quark to hadron polarization in the hadronization process is not yet well understood~\cite{Ayala1, Ayala2}, our findings show that if the hadronization mechanism preserves a memory of the constituent quark polarization, a difference between hadron and antihadron polarization may be expected, particularly at large values of $\mu$.

\section*{Acknowledgements}

Support for this work has been received by Consejo Nacional de Ciencia y Tecnolog\'ia grant number 256494 and by UNAM-DGAPA-PAPIIT grant number IG100219.

\end{document}